\documentclass[preprint,11pt]{elsarticle}
\usepackage[english]{babel}
\usepackage{graphicx,color}
\usepackage{amsmath}
\usepackage{amssymb}
\usepackage[utf8]{inputenc}

\graphicspath{{./figures/}}

\def\gtrsim{\mathbin{\;\raise1pt\hbox{$>$}\kern-8pt\lower3pt\hbox{$\sim$}\;}}

\begin{document}
\begin{frontmatter}
\title{Binary Neutron Star Merger Remnants as Sources of Cosmic Rays Below the ``Ankle''}

\author[add1]{X.~Rodrigues}
\author[add1]{D.~Biehl}
\author[add1]{D.~Boncioli}
\author[add1]{A.~M.~Taylor}

\address[add1]{Deutsches Elektronen-Synchrotron (DESY), Platanenallee 6, D-15738 Zeuthen, Germany}

\date{}


\begin{abstract}
We investigate non-thermal electron and nuclei energy losses 
within the binary neutron star merger remnant produced by the  
event GW170817. The lack of a cooling feature within the detected 
synchrotron emission from the source is used to constrain the magnetic field 
at the mG level, assuming that this emission is electron 
synchrotron in origin, and that the accelerated spectrum in the electrons follows 
the form $dN/dE_e \propto E_e^{-2}$. The level of subsequent gamma-ray 
emission from the source is demonstrated to provide a further 
constraint on the source magnetic field strength.
We also put forward alternative strong ($\sim$G) magnetic field scenarios able to
support this emission. For such stronger fields,
the photo-disintegration of non-thermal nuclei within the source is considered, 
and a bottleneck period of $\sim$5-30~days is found when this
process peaks. We find that this class of source is in principle 
able to support the population of cosmic rays detected at Earth below
the ``ankle''.
\end{abstract}


\end{frontmatter}

\section{Introduction}
\label{intro}
The gravitational wave event detected on 17~August, 2017 \cite{TheLIGOScientific:2017qsa} originated from the first discovered binary neutron star (BNS) merger event GW170817. Immediately 
following this event (within $2$~s), prompt short GRB emission was detected by both Fermi-GBM and the 
INTEGRAL-SPI detectors \cite{Monitor:2017mdv,Savchenko:2017ffs}. Subsequently, a relatively 
bright electromagnetic counterpart, EM170817, was discovered in optical bands
\cite{Arcavi:2017fuz,Coulter:2017wya,Lipunov:2017dwd,Soares-Santos:2017lru,Valenti:2017ngx}. 
This optical counterpart is associated to emission from the remnant. An estimation on the distance to 
the remnant from the GW signal was found to be consistent with that of the nearby
galaxy NGC~4993 \cite{Hjorth:2017yza} ($\sim $40~Mpc), connecting the GW source with this host galaxy.

Subsequent non-thermal observations of the remnant in both radio and X-ray bands 
\cite{Hallinan:2017woc,Alexander:2017aly,Margutti:2017cjl,Ruan:2017bha} indicated that efficient in situ particle acceleration is taking 
place within the source. This emission was observed to brighten as a power-law in time for a 
timescale of $\sim$160~days, with the apparent recent onset of a dimming of the source
\cite{DAvanzo:2018zyz,Nynka:2018vup}.

The outflowing kinetic energy (KE) released in the blast wave, estimated from gravitational
binding energy considerations, may be of the order of $10^{50-52}$~erg. An approximate value of total KE in the outflow 
of $10^{51}$~erg would be consistent with an estimated mass of $10^{-2}~M_{\odot}$  of material being 
ejected with velocity $\beta=0.2$. The total number of electrons in the outflow is then $N_e=10^{55}$, which means at late times ($t \approx 100$~days), the estimated 
average density of material in the outflow is $\bar{n}\approx 2\times10^{4}~{\rm cm}^{-3}$. 
Depending on the rate at which the blast wave entrains material, the non-thermal 
energy density in the relativistic particles, and magnetic field, may eventually rise up into 
closer balance with the outflowing ram pressure ($p_{\rm ram}\sim 10^{11}~{\rm eV~cm}^{-3}$) on the Sedov-phase 
timescale.

For such a massive compact outflow, two important timescales are worth noting. Firstly, the
opacity of the source to its own radiation is
\begin{equation}
\tau_{e\gamma} \,=\, \frac{N_{e} \sigma_T R}{V} \,\approx\, \frac{N_{e} \, \sigma_T}{4 (\beta c t)^2} \,\approx\, \left(\frac{t}{2~\mathrm{days}}\right)^{-2}.
\end{equation}
Thus, the early compact source would be expected to have undergone a transition from being optically 
thick to thin on a timescale of $t\sim 2$~days (see also \cite{Piro:2017ayh}).

Secondly, adopting an ejecta mass of $10^{-2}M_{\odot}$, the Sedov phase occurs when 
$\bar {n}  =M/m_{p}(\beta c t)^{3}$, giving
\begin{eqnarray}
t=0.6~\frac{0.2}{\beta}\left(\frac{10^{4}~{\rm cm}^{-3}}{\bar{n} }\right)^{1/3}~{\rm yr},
\end{eqnarray} 
Thus, depending on the mean ambient density that the remnant expands into ($\bar{n}$), 
the non-thermal brightening of the source would be expected to cease on a $\sim 100$~day timescale.

The case for late-time acceleration of particles in outflows from BNS merger events 
was anticipated in \cite{Takami:2013rza}.
These authors motivated the magnetic fields to be $\sim$mG in strength, an outflow mass of $\sim 10^{-2}~M_{\odot}$ and a non-relativistic shock speed, and inferred a maximum proton energy of 
$\sim 10^{17}$~eV for the case of particle acceleration up to the deceleration timescale 
$t_{\rm dec}\approx 1$~yr, after which the outflows velocity starts to decelerate (presumed to reach 
the Sedov phase). Indeed, the mildly relativistic aspect of these outflows motivates them
as interesting cosmic ray (CR) acceleration sites on theoretical grounds \cite{Bell:2017zzx}.
Thus, these sources can be naturally motivated to be potential accelerators of extragalactic CRs
 in the ``knee'' to ``ankle'' regions of the spectrum.

Here we focus on both electron and nuclei acceleration within the source, considering the
subsequent energy losses and emission. We further assess the potential interactions that 
accelerated nuclei may undergo within the source's radiation fields.
In Sec.~\ref{B-field_Constraints}, constraints are placed on the source magnetic field under the
assumption that electrons accelerated by the source possess a  Fermi shock acceleration type spectrum. 
In Sec.~\ref{Electron_Losses}, consideration is made on the number of non-thermal 
synchrotron-emitting electrons required to power the source at its observed brightness level, 
dependent on the source magnetic field strength, with both ``low'' and ``high'' magnetic field 
strength values being considered. The level of the inverse 
Compton (synchrotron self-Compton, SSC) emission produced for the various magnetic field strengths considered is also addressed.
In Sec.~\ref{Nuclei_Acceleration}, the subsequent interaction of non-thermal nuclei accelerated by the source,
with the intense thermal radiation field observed on early (multi-day) timescales after the merger,
are determined. The conclusions are made in Sec.~\ref{Conclusion}.
Throughout this discussion, we assume that the non-thermal emission produced on days to hundreds
of day timescales is emitted isotropically by the source.

\section{Magnetic Field Constraints}
\label{B-field_Constraints}

Photospheric velocity measurements of the remnant EM170817 provide evidence supporting
the presence of an outflow with non-relativistic speeds ($\beta\lesssim 0.2$) \cite{Piro:2017ayh}. Furthermore, 
this outflow has been observed to give rise to non-thermal emission whose brightness has been growing 
with time at both radio ($\sim 0.3-6\times 10^{-5}$~eV) and X-ray ($\sim 0.3-8\times 10^{3}$~eV) energies 
\cite{Mooley:2017enz,Ruan:2017bha}. This emission, assumed synchrotron in 
origin, may be emitted by an electron population accelerated to a spectrum 
$dN/dE_{e}\propto E_{e}^{-2}$. The apparent lack of a 
cooling feature in the observed spectrum, between radio and X-ray energies, can be used to constrain the synchrotron cooling rate within the source. Throughout this work we will assume $\beta=0.2$ for the the calculation of numeric results, while noting that the qualitative conclusions do not depend on the exact  value assumed for the outflow speed.


\textit{Dynamical time limited acceleration.} 
The above-mentioned constraint on the minimum magnetic field strength can also be placed from a 
consideration of the acceleration timescale,
\begin{eqnarray}
t_{\rm acc}=\left(\frac{1}{\beta^{2}}\right) t_{\rm Lar},
\end{eqnarray}
where $t_{\rm Lar}$ is the Larmor time. For the requirement that $t_{\rm acc}<t_{\rm dyn}$, this converts 
to $t_{\rm Lar}<4\times10^{-2}~t_{\rm dyn}$ (adopting $\beta=0.2$).

\textit{Cooling time limited acceleration.} 
The lack of a cooling feature in the observed synchrotron spectrum can be used to place a constraint
on the maximum magnetic field giving rise to the observed synchrotron emission.
The synchrotron cooling time of mono-energetic electrons with Larmor radius giving 
rise to synchrotron photons with characteristic energy $E_{\gamma}^{\rm syn}$ is
\begin{align}
t_{e}^{\rm syn} &= \frac{9}{8\pi\alpha}\left(\frac{m_{e}}{E_{\gamma}^{\rm syn}}\right)t_{\rm Lar} \nonumber \\
&= 2\times 10^{3}\left(\frac{10~{\rm keV}}{E_{\gamma}^{\rm syn}}\right)t_{\rm Lar}.
\label{eq:synchrotron_timescale}
\end{align}
Assuming the electron spectrum is produced via Fermi non-relativistic shock acceleration, giving rise to $dN/dE_{e}\propto E_{e}^{-2}$, the lack of a 
cooling feature in the observed synchrotron spectrum 
constrains the cooling time of the electrons producing this emission, 
$t_{\rm dyn}<t_{e}^{\rm syn}(E_{\gamma}^{\rm syn}=10~{\rm keV})$, leading to the
constraint $t_{\rm Lar}>4\times 10^{-4}~t_{\rm dyn}$. 

Thus, overall, a dual constraint on the Larmor period of the particles in the system is found
of, 
\begin{eqnarray}
4\times 10^{-4}<t_{\rm Lar}/t_{\rm dyn}<4\times10^{-2},
\label{constraint_tlar}
\end{eqnarray}
where $t_{\rm dyn}\sim 100~{\rm days}\sim 9\times 10^{6}$~s. As a reference, it is useful to note 
that a 100~TeV electron in a mG magnetic field has a Larmor period of $\sim 6\times 10^{4}$~s.

The Larmor time relates to the synchrotron photon energy emitted ($E_{\gamma}^{\rm syn}$) 
and the emitting electron energy ($E_{e}$) via the relation,
\begin{eqnarray}
t_{\rm Lar} = \left(\frac{E_{e}}{m_{e}}\right)^{3}\left(\frac{\hbar}{E_{\gamma}^{\rm syn}}\right).
\label{tlar}
\end{eqnarray}
The constraint on $t_{\rm Lar}/t_{\rm dyn}$ in expression~\ref{constraint_tlar}, substituted into Eq.~\ref{tlar} subsequently provides a constraint on the
maximum electron energy, given the observation of hard ($\sim10~{\rm keV}$) X-ray emission (cf.~Sec.~\ref{Inverse_Compton}). This in turn constrains the magnetic field strength in the source.
The dynamical time limit, for which $t_{\rm Lar}=4\times10^{-2}~t_{\rm dyn}$, places the acceleration energy 
scale at $\sim 10^{14}$~eV, and a corresponding magnetic field strength of $\sim 0.03~{\rm mG}$. Conversely,
the cooling time limit, for which $t_{\rm Lar}=4\times 10^{-4}~t_{\rm dyn}$, places these scales at 
$\sim 9\times 10^{12}$~eV and the magnetic field strength at a value of 2~mG. Therefore, the 
overall dual constraint on the magnetic field in the system is:
\begin{eqnarray}
0.03~{\rm mG}<B<2~{\rm mG}.
\label{eq:b_constraints}
\end{eqnarray}

In Sec.~\ref{Inverse_Compton} we will argue that gamma-ray observations can provide a stronger constraint on the minimum magnetic field strength. On the other hand, if assuming the maximum value of $B=2~{\rm mG}$ throughout the age of remnant, the corresponding maximum energy for protons (assuming no energy losses) is
\begin{eqnarray}
E_{p}^{\rm max}
&\approx& 6\times 10^{15}\left(\frac{\beta}{0.2}\right)^2\left(\frac{t_{\rm dyn}}{100~{\rm days}}\right)\left(\frac{B}{2~{\rm mG}}\right)~{\rm eV},
\end{eqnarray}
obtained by balancing $t_{\rm acc}$ with $t_{\rm dyn}$. While in Sec.~\ref{high_B-field} we will argue for the possibility of a different scenario with strong $\mathcal{O}$(G) magnetic fields being present within the source, it is worth noting that already within the above constraints, this source type is capable of accelerating CRs to energies beyond the knee and below the ankle. Such an energy region in the CR spectrum is interesting due to the indications for the onset an additional source component~\cite{hillas2004a}, which may help bridge the energy gap between the iron knee \cite{Apel:2011mi} and the ankle. Furthermore, constraints on the dipole anisotropy at such energies strongly motivate the idea that this additional component is extragalactic in origin \cite{Giacinti:2011ww,Abreu:2012ybu}. The new BNS mergers source class appears as an interesting new candidate due to its proven capability to operate as an efficient particle accelerator.

\section{Non-Thermal Electron Losses}
\label{Electron_Losses}

In actuality, two alternative (extreme) scenarios exist that can explain the observed luminosity of 
the synchrotron emission. With a low magnetic field, like that obtained above, an electron energy density 
is needed that far surpasses the magnetic energy density (see discussion in Sec.~\ref{low_B-field}). 
However, an alternative scenario is
that a high magnetic field is present (to boost synchrotron production), although this violates the 
constraint derived previously from the lack of a cooling feature. This violation can be negated, however, 
in two possible ways. Firstly, if the 
observed synchrotron spectrum is actually all produced by cooled electrons, injected by the source with a 
much harder acceleration spectrum than that produced by Fermi acceleration. Or secondly, if the emitted radiation originates in fresh electrons, continuously picked up at the edge of the outflow and accelerated at the shock front. In the 
following we discuss possible parameter sets representative of either extreme magnetic field case.

\subsection{Slow acceleration scenario}
\label{low_B-field}
In this scenario, we consider a weak, uniform magnetic field present within the remnant. As an example, we consider the value $B=0.2~{\rm mG}$, which lies within the range given in Eq.~\ref{eq:b_constraints}.

The peak of the synchrotron power emitted by an electron distribution is dominated by the 
highest-energy electrons accelerated. Assuming an electron synchrotron origin of the observed X-ray 
emission, it is dominated by electrons of energy

\begin{equation}
  E_e^{\rm X-ray} \,=\, 30 \left(\frac{E_\gamma}{10~{\rm keV}} \right)^{1/2} \left(\frac{B}{2~\mathrm{mG}} \right)^{-1/2}~{\rm TeV}.
  \label{eq:xray_energy}
\end{equation}

In this monochromatic approximation, the total X-ray luminosity emitted by a population of electrons is

\begin{equation}
  L_{\rm syn}^{\rm X-ray} \,\approx \, c \, \sigma_T \, (E_{e}^{\rm X-ray}/m_{e}c^2)^2 \, u_B \, N_e^{\rm X-ray},
  \label{eq:lum_synch}
\end{equation} 
where $N_{e}^{\rm X-ray}$ is the number of X-ray-emitting electrons. As a reference for the synchrotron 
luminosity we rely on observations around 110 days after the event~\cite{Margutti:2017cjl}. Using 
an approximate value of $L_{\rm syn}^{\rm X-ray}=4\times10^{39}~{\rm erg~s}^{-1}$ at $10~{\rm keV}$, 
Eq.~\ref{eq:lum_synch} yields $N_e^{\rm X-ray}=3\times10^{46}$. Since the number of non-thermal 
electrons goes as $E_{e}\frac{dN}{dE_{e}}\propto E_{e}^{-1}$, the total number of accelerated 
electrons is dominated by the lowest-energy, radio-emitting electrons, whose number is given by 
\begin{equation}
  N_e^{\rm radio} = N_e^{\rm X-ray} \frac{E_e^{\rm X-ray}}{E_e^{\rm  radio}}.
  \label{eq:n_min}
\end{equation} 

This yields a number of radio-emitting electrons ($E_\gamma=10^{-6}~{\rm eV}$) of $3\times10^{51}$. 
In this scenario, it is important to note that if the electron population were to extend to energies lower than $E_{e}^{\rm radio}\approx300~{\rm MeV}$,
 the total electron population could reach a number close to the maximum possible number of
swept-up and injected electrons at the Sedov phase (some $N_e^{\rm total}=10^{55}$).
This is particularly so bearing in mind that only 1\% of the non-thermal particle population within sources are thought to be 
electrons \cite{Burbidge:1959}.
It is also informative to note that in this scenario the electron energy density is much higher than the magnetic energy 
density, $u_e\approx10^7 u_B$, assuming a volume of $7\times10^{50}~{\rm cm}^3$ corresponding to a sphere expanding 
with velocity $\beta=0.2$ after 110 days.

\begin{figure}[tbp]
  \includegraphics[width=\textwidth, angle=0]{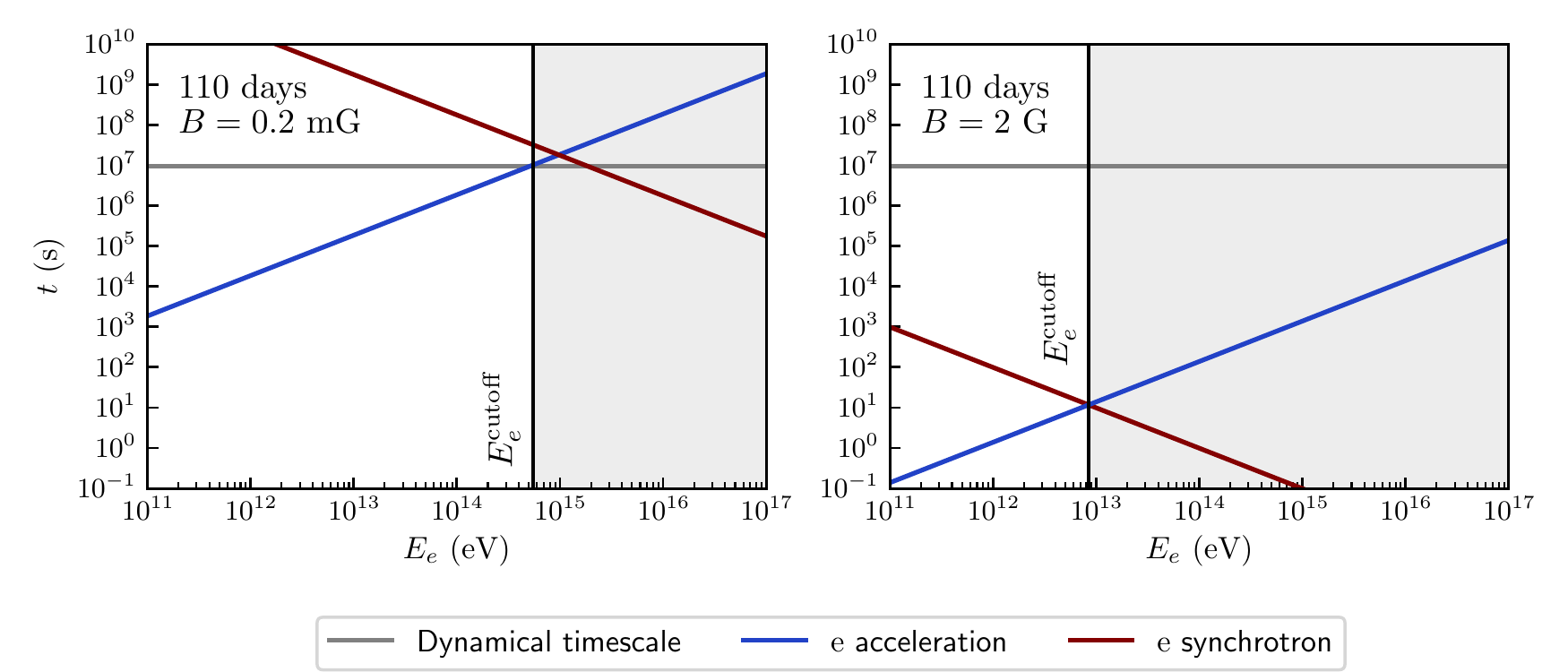}
  \caption{Relevant timescales for a magnetic field strength of $B=0.2~{\rm mG}$ (left) and $B=2~{\rm G}$ (right). The dynamical timescale (gray line) is given by the expansion time of the remnant, which determines the efficiency of electron acceleration (blue) and cooling (red). In the case of a weak magnetic field, a cutoff on the electron distribution is expected at around 600 TeV, when acceleration is no longer efficient given the age of the remnant. On the other hand, in the strong magnetic field case a cooling cutoff is expected at $8~{\rm TeV}$, when synchrotron cooling dominates over acceleration.}
  \label{fig:timescales}
\end{figure}

Finally, note that the maximum energy that may be achieved by electrons in the source may be higher 
than that which dominates X-ray production. In fact, as shown in the left panel in Fig.~\ref{fig:timescales}, in a $2~{\rm mG}$ field electrons can be accelerated in the source up to $E_e^{\rm max}=700~{\rm TeV}$ by the 110 day timescale. At these energies, acceleration is limited by the age of the remnant and a cutoff is expected. A cooling break is not expected, since synchrotron emission (blue line in  Fig.~\ref{fig:timescales}) is not efficient in the relevant energy range.

\subsection{Fast acceleration scenario}
\label{high_B-field}

A somewhat opposite scenario is possible involving a stronger magnetic field strength. In fact, a scenario with strong magnetic fields would be supported by CR-driven magnetic field amplification expected in environments with strong CR fluxes~\cite{bell2004}. This instability may in principle drive the magnetic field strength up to a saturation value of as much as $2~{\rm G}~(M/10^{-2}~M_{\odot})^{1/2}(t/{\rm 100~days})^{-3/2}$ (assuming $p_{\rm ram}\sim u_{\rm CR}$), an estimate that is independent of the outflow speed.
Furthermore, such strong magnetic fields are at the limit allowed in order that the shock remain super-Alfvénic, beyond which efficient diffusive shock acceleration of CRs would not be possible. For a magnetic field strength of 2 G, the shock speed $\beta=0.2$ is still higher than the Alfvén velocity of the plasma, $\beta_A=0.15(B/2~{\rm G})$.

One possibility to explain observations given a strong magnetic field scenario is that the electron population is accelerated to a harder spectrum by a mechanism other than diffusive shock acceleration, such as stochastic acceleration in the regime with acceleration timescales out-competing escape timescales \cite{Jones1994}. 
In such strong magnetic field case, with the radiation zone different from the acceleration zone, the entire non-thermal electron population would be cooled to a $E^{-2}$ spectrum, yielding the observed power law synchrotron emission.

Alternatively, a strong magnetic field in the source may be present if the emitted radiation continuously originates from electrons accelerated near the edge of the outflow. In fact, as the remnant expands, the number of electrons picked up grows with the volume, while the magnetic field strength may conceivably decrease linearly with time, as inferred for other fast moving outflows such as the supernova SN 1993J~\cite{Fransson:1998ma,Tatischeff:2009kh}. This would imply that synchrotron emission is dominated by freshly accelerated electrons, which dominate in number, thereby relaxing the constraint on the synchrotron cooling efficiency of the source.

Regardless of the origin of the synchrotron emission, in a high magnetic field scenario the observation of radio emission can be used to constrain the efficiency of synchrotron self-absorption in the remnant at those energies (\textit{cf.} Sec.~\ref{Inverse_Compton}), which in turn can be used to constrain the magnetic field to a maximum of 10~G.



For such strong magnetic fields the observed X-ray luminosity is produced by lower energy electrons than in the weak magnetic field 
scenario. Adopting the value $B=2~{\rm G}$, we have $E_e^{\rm X-ray}\approx300~{\rm GeV}$ (\textit{cf.}~Eq.~\ref{eq:xray_energy}). To account for the observed X-ray luminosity, the necessary number of high-energy electrons is $N_e^{\rm X-ray}=3\times10^{42}$, which by 
Eq.~\ref{eq:n_min} yields a number of radio-emitting electrons of $3\times10^{47}$. This value is much lower 
than that derived for the scenario discussed in the previous
section, which means that even if the electron population extends to lower energies their number cannot approach the total amount of
swept-up electrons. In this high magnetic field scenario, however, the energy density of the non-thermal electrons 
and magnetic field are related by $u_e\approx 10^{-7}u_B$.

Finally, as shown in the right panel of Fig.~\ref{fig:timescales}, the maximum electron energy allowed in this 
case is only $8~{\rm TeV}$. At that energy synchrotron cooling dominates over acceleration at higher energies, and a cooling cutoff is expected~\cite{Aharonian:2000pv}, a characteristic that is not expected in the low magnetic field scenario.

\subsection{Inverse Compton Emission}
\label{Inverse_Compton}

A population of non-thermal particles embedded in ambient radiation fields invariably give
rise to subsequent inverse Compton emission. Assuming that the observed synchrotron radiation 
dominates the ambient radiation field seen by the non-thermal particle population, the inverse
Compton emission produced will be dominated by synchrotron self-Compton scattering (SSC). 
For cases in which $u_{e}\gg u_{B}$, considerable
SSC emission can be expected \cite{Fransson:2004kw}.

In order to determine the inverse Compton emission at different times, a prescription
for the radiation field evolutions must be adopted. The non-thermal radiation field is 
normalized by fixing the X-ray luminosity, obtained from observations at 9, 15 and 110~days 
after the merger \cite{Margutti:2017cjl,Troja:2017nqp,Ruan:2017bha}, and subsequently extrapolating it back in time 
assuming a continuation of the inferred power-law evolution of the form,
\begin{eqnarray}
L_{\rm X-ray} = 2\times 10^{39}~(t/110~{\rm days})^{0.6}~{\rm erg~s}^{-1}.
\label{non_therm_evolv}
\end{eqnarray}

At early times, $t< 15~{\rm d}$, a bump is observed in the optical 
range of the SED, with a spectral shape characteristic of thermal emission 
\cite{Margutti:2017cjl}. This thermal bump is seen to decrease between 10 and 74~days after the 
event~\cite{Villar:2018mqw}. Based on these observations, we model the evolution of this thermal
luminosity as
\begin{align}
L_{\rm th} &=  4\times 10^{40}~{\rm erg~s}^{-1} && (t<~7~ {\rm days})\nonumber\\
  &=  4\times 10^{40}~(t/7~{\rm days})^{-2.3}~{\rm erg~s}^{-1} && (t>7~ {\rm days}).
\label{therm_evolv}
\end{align}

\begin{figure}[t!]
  \includegraphics[width=\textwidth, angle=0]{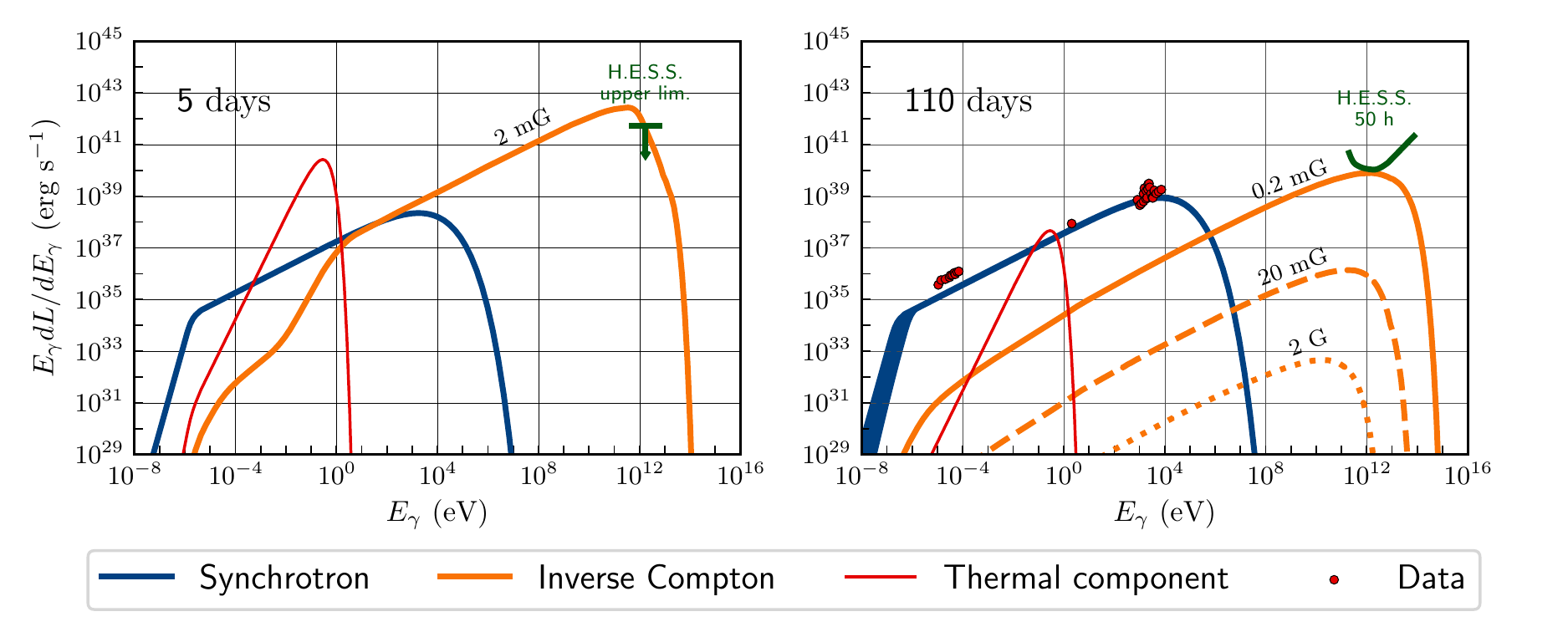}
  \caption{Luminosity spectrum produced by electron synchrotron emission (blue, including synchrotron self-absorption), and inverse-Compton scattering (yellow), at five days (left) and 110 days (right) after the merger event. For the 110 day case we show the results for three values of $B$-field strength. In red we show optical~\cite{Lyman:2018qjg}, radio and X-ray data~\cite{Margutti:2018xqd} at 110 days. In the left panel we show in green the limit on the TeV luminosity at 5 days~\cite{Abdalla:2017mtd} and on the right panel we show the H.E.S.S. sensitivity, adopted from~\cite{Holler:2015uca}, which translates into an upper limit on the source gamma-ray luminosity.}
  \label{fig:ic_sed}
\end{figure}

In Fig.~\ref{fig:ic_sed} we show the resultant broadband SED produced by the remnant at 5 and 
110~days after the merger. Note that the low-energy cut-on of the synchrotron spectra (blue curves) is given by synchrotron 
self-absorption, which is included in our emission model, following the prescription provided in 
\cite{Longair:1994wu}. 

In the left panel we show the case where a 2~mG magnetic field is 
present in the outflow at 5~days after the merger event. At these early times in the remnant, the
thermal radiation field (red curve) provides the dominant target for inverse Compton emission (yellow curve).
We also show the H.E.S.S. upper limit, in the range 0.5-6~TeV, at 5.2~days \cite{Abdalla:2017mtd}. 
The magnetic field at these early timescales has been adopted sufficiently high so as to 
ensure that the inverse Compton emission does not overshoot the H.E.S.S. upper limit.
For this case, a sharp cutoff is introduced into the inverse Compton spectrum, due
to pair production on the thermal radiation field, which we include in 
our emission model following the prescription of \cite{Gould:1967zzb}. 
We note that following pair production within the source, subsequent cascade 
development and emission would result, leading to the redistribution of the spectrum at lower 
energies. However, for the purposes of comparing the inverse Compton flux to the H.E.S.S. 
observation upper limit, this additional lower-energy component may be neglected.
The considerable level of inverse Compton emission found for the case of low magnetic field
demonstrates that at later times, once the thermal bump has reduced sufficiently, a lower limit on 
the magnetic field strength may be placed at late timescales, by follow-up TeV observations 
of the remnant. 


In the right panel of Fig.~\ref{fig:ic_sed} we show the resultant broadband SED produced by
the remnant at 110 days, normalized to the 
observed X-ray luminosity, and we show the resultant inverse Compton emission for three values of magnetic field strength.
At this later time in the remnant, the synchrotron radiation field provides the dominant
target for inverse Compton emission.
The red points in this plot show radio, optical, and X-ray data points taken at this time 
\cite{Margutti:2018xqd,Lyman:2018qjg}.

For the weak magnetic field scenario result ($B=0.2~{\rm mG}$, right panel of
Fig.~\ref{fig:ic_sed}), the inverse Compton luminosity  
dominance is highest and the predicted gamma-ray luminosity is $10^{40}~{\rm erg~s}^{-1}$, 
peaking at $E_\gamma=1~{\rm TeV}$. We can see that this emission is at the level of the H.E.S.S. 50 h sensitivity (green curve \cite{Holler:2015uca}), which indicates that this instrument has the required sensitivity to set a lower limit on the magnetic field strength in the source.
In contrast, for higher magnetic field strengths, we see that the low 
electron density yields a gamma-ray power of only 
$3\times10^{32}~{\rm erg~s}^{-1}$ with a $10~{\rm GeV}$ peak. This result demonstrates that the 
potential probing of the Compton peak by TeV instruments can provide fresh insights to
discriminate between the different magnetic field strengths considered. 

\section{Acceleration of cosmic-ray nuclei}
\label{Nuclei_Acceleration}

\subsection{Acceleration to energies between the knee and the ankle}
\label{Acceleration_Below_Ankle}

The determination of the interaction processes which dictate the maximum energy of the accelerated 
nuclei depends on the magnetic field strength adopted. Since the low magnetic field value discussed 
in Sec.~\ref{low_B-field} would increase the acceleration time, acceleration up to high energies 
and the onset of nuclear photo-disintegration would not be possible. For the purpose of 
investigating these interactions, we adopt in the following the extreme magnetic field scenario 
discussed in Sec.~\ref{high_B-field}.

\begin{figure*}[t!]
\centering
\includegraphics[width=.7\textwidth]{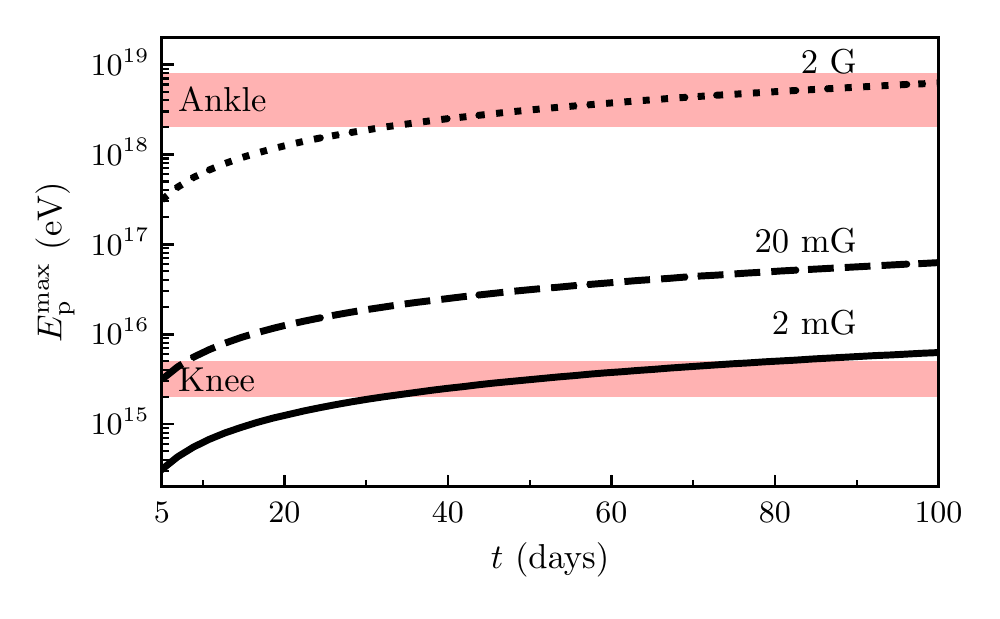}
\caption{Maximum energy achieved by protons accelerated in the remnant as a function of time since the merger, for three different magnetic field strengths. The age of the remnant is always the limiting factor to the maximum proton energy, as photo-meson production is never efficient. As throughout this work, an expansion velocity of $\beta=0.2$ has been assumed.}
\label{fig:cr_emax}
\end{figure*}

As shown in the following section, photo-hadronic cooling of protons in the source is never efficient, even with the bright optical emission at early days. Instead, the maximum energy achieved by protons is always limited by the age of the remnant. In Fig.~\ref{fig:cr_emax} we show the evolution of the maximum proton energy with time, for three constant values of magnetic field strength. We can see that in a weak magnetic field scneario, the source can only accelerate cosmic rays above the knee from around 80 days after the merger onwards. On the other hand, in the case of a strong magnetic field scenario, late-time acceleration of protons is possible up to the ankle, as shown in Fig.~\ref{fig:cr_emax} for the maximum magnetic field strength (2~G) we consider.

\subsection{Energy losses at the ankle}
\label{Losses_Ankle}

Like electrons, nuclei accelerated by the source will also interact with the target photons present within it. In fact, the appreciable attenuation at early times  ($<10$~days) of TeV photons in the source found in Sec.~\ref{Inverse_Compton} gives reason to expect also considerable photo-disintegration in the source on these timescales \cite{Neronov:2007mh,Murase:2010va,Aharonian:2010te}. 

\begin{figure*}[t!]
\centering
\includegraphics[width=\textwidth]{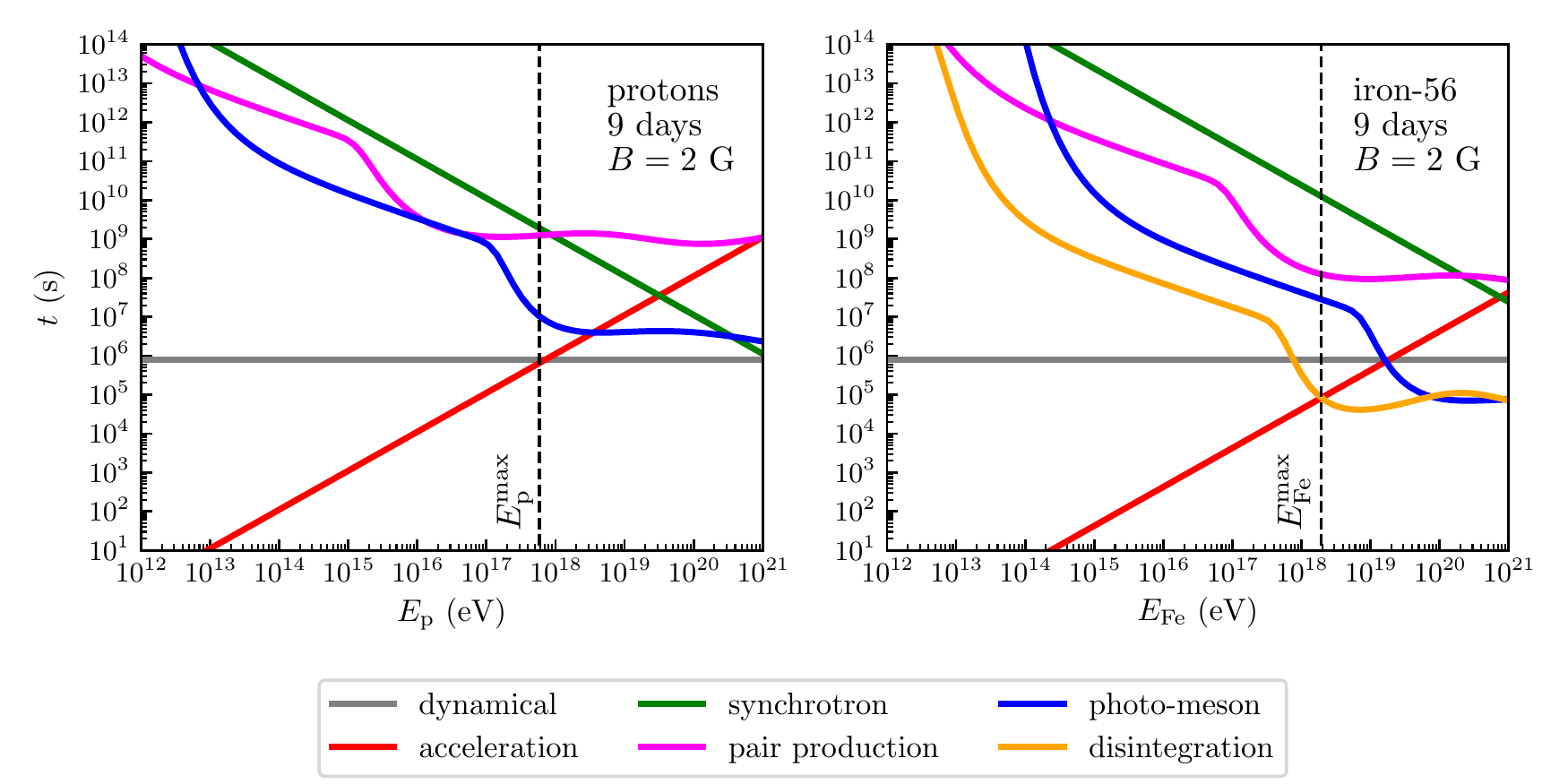}
\caption{Interaction times for protons (left) and iron-56 nuclei (right) as a function of the particle energy, 9 days after the merger, in the high magnetic field scenario. Note that for the the synchrotron and pair production processes we show the energy loss timescale.}
\label{fig:iron_rates}
\end{figure*}

In Fig.~\ref{fig:iron_rates} we show the interaction timescales of the different processes at work 
for both protons (left panel) and iron-56 nuclei (right panel), obtained using the \textsc{NeuCosmA} code, 
which has been previously developed for the study of Gamma-Ray Bursts \cite{Biehl:2017zlw} and blazars 
\cite{Rodrigues:2017fmu}. We take as reference the time interval of 9 days after the merger, when photo-hadronic interactions are most efficient, as discussed later in this section. In the case of protons, we see that acceleration (red line) is always limited by the age of the remnant, which corresponds to the dynamical timescale of the system, shown in the gray line. On the other hand, the source is seen to be optically thin to photo-meson production (blue curve). We also plot the pair production loss time (magenta curve), showing that although it is always a sub-dominant energy 
loss process for hadrons, it will contribute appreciably to the electron population.

On the right panel of Fig.~\ref{fig:iron_rates}, we can see that at the 9 day timescale the bright optical radiation can efficiently photo-disintegrate iron-56 nuclei, as shown by the yellow curve, achieving an optical thickness (i.e. the ratio between the dynamical and photo-disintegration timescales) of about 10 at 2~EeV. At this energy, iron nuclei cannot be efficiently accelerated, as photo-disintegration becomes the dominant process and the nuclei instead disintegrate into lighter isotopes. At later times, however, the thermal photon luminosity decreases (see \ref{therm_evolv}), making photo-disintegration less efficient, and the maximum energy becomes limited only by the age of the remnant, as in the case of protons.

\begin{figure*}[t!]
\centering
\includegraphics[width=.95\textwidth]{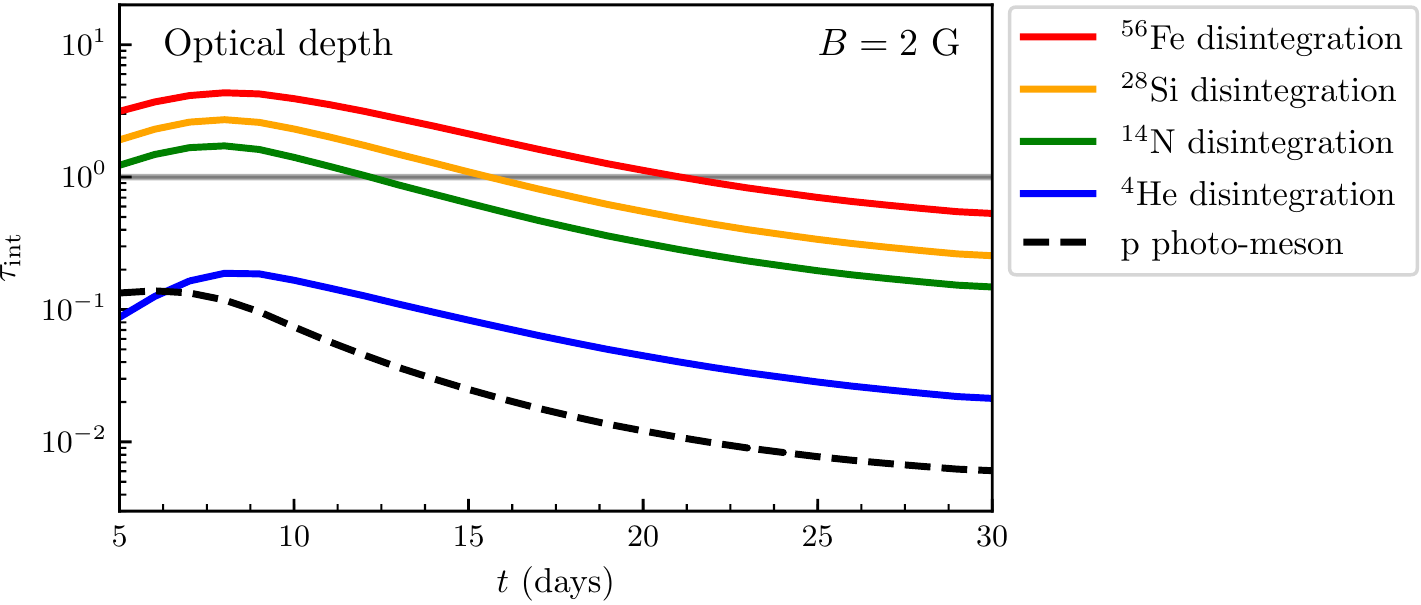}
\caption{Optical depth of the remnant to photo-disintegration of different isotopes (solid curves) and photo-meson production by protons (dashed curve) as a function of the time since the merger, in the high magnetic field scenario. The horizontal gray line represents the transition from optically thin to thick (see main text).}
\label{fig:disintegration}
\end{figure*}

In Fig.~\ref{fig:disintegration} we show the temporal evolution of the optical depth of the source 
to photo-disintegration of different isotopes, as well as photo-meson production by protons. Here, the 
optical depth is defined as $\tau_{\rm int} = \left.t_{\rm dyn}/t_{\rm int}\right |_{\rm E_{\rm max}}$, where 
$t_\text{dyn}$ is the dynamical timescale and $t_\text{int}$ the interaction timescale, evaluated 
at the maximum energy of the CR (see vertical dashed line in Fig.~\ref{fig:iron_rates}).
This provides a measure for the fraction of accelerated CRs that escape the source. 
The optimal time for nuclear photo-disintegration is found to 
be reached within the first weeks after the merger. This happens because of the competing 
processes within the expanding object, namely the rise of the maximum particle energies 
achievable and the decrease of the thermal photon field density.

Assuming continuous injection, the late time acceleration of protons can bring them to energies up to the ``ankle''. 
On the other hand, the composition of CRs eventually escaping the source can be quite different depending on the period of efficient photo-disintegration, on the primary isotope, and on the escape mechanism assumed. As the present study is held very general, we do not make an effort to calculate the resultant late-time accelerated nuclei spectra. 

\subsection{Cosmic rays from the merger population}
\label{Population}

With some level of photo-disintegration expected within the source during 
early times, and very high energies becoming within reach at later times, 
we finally turn our attention 
to the ensemble population of such sources. We implicitly assume here that the event we discuss 
throughout this work is representative of a population of identical sources which could accelerate 
CR nuclei. In this scenario, there should be a number of electromagnetic counterparts to this 
population, with the event EM170817 being the only one detected so far.

To support a CR spectrum at Earth with the abundance observed \cite{Gaisser:2013bla}, a 
local emissivity $\mathcal{L}_0$ of their sources is required. Assuming a CR luminosity density of 
$\mathcal{L}_0\sim 4\times 10^{44} ~{\rm erg~yr}^{-1}~{\rm Mpc}^{-3}$  \cite{Waxman:1995dg}, and
adopting a specific local rate for these sources, one can infer the required energy released
per event. Considering a local rate of BNS mergers of $1540^{+3200}_{-1220}~\mathrm{Gpc^{-3}~yr^{-1}}$~\cite{TheLIGOScientific:2017qsa}, we estimate the required energy input into CRs in each merger event of $E_{\mathrm{CR}}\approx2\times 10^{50}~\mathrm{erg}$. This estimate is roughly consistent with a fraction of $\sim$10\% of the released outflowing KE given in Sec.~\ref{intro}.

Using the previous value for the total energy output in CRs, 
we can also estimate the total neutrino fluence produced by the source. 
From Figs.~\ref{fig:iron_rates} and \ref{fig:disintegration} we see that the optical thickness of the source at the 9 day timescale is about $\tau_{{\rm p}\gamma}=0.1$, while for later times it diminishes due to the dimming of the thermal luminosity (Eq.~\ref{therm_evolv}) and the overall reduction in photon density due to expansion. The average optical thickness from 5 to 160 days is then approximately $\langle\tau_{{\rm p}\gamma}\rangle=5\times10^{-3}$.
The resulting estimate of the neutrino fluence from the source is 
$E_{\mathrm{CR}} K_{p\gamma}\langle\tau_{{\rm p}\gamma}\rangle/(4\pi d^{2}) \approx 5\times 10^{-4}~{\rm GeV~cm}^{-2}$, where $K_{p\gamma}\approx 0.15$ describes 
the fraction of energy taken by a pion in each photo-meson production interaction and $d=40~{\rm Mpc}$ is the distance to the source. Note that the normalization factor for this estimate is provided by the energy input 
in CRs deduced above, which carries considerable uncertainty, mainly due to the 
present large uncertainty on the source rate. However, a simple comparison with the present upper limits 
for the neutrino flux from EM170817 \cite{ANTARES:2017bia} indicates that the level of this flux
would be challenging to reach by present instruments like IceCube. For comparison, the case of neutrino 
emission from the BNS merger pulsar remnant 
was considered in \cite{Fang:2017tla,Kimura:2017kan}. In both cases, rather increased neutrino fluxes 
were found to be expected compared to those determined here, which is due to the faster acceleration process at play in 
the source and the differing source environment.

\section{Conclusion}
\label{Conclusion}

Under the premise that the observed non-thermal emission from the remnant is isotropic,
we consider the non-thermal particle energy losses within the fast-moving 
remnant outflow associated with EM170817 . Assuming that this emission is electron in
origin, the consideration of the lack of a cooling
break in the synchrotron emission from this remnant, whose age is accurately known, 
allowed a constraint to be made on the strength of the source magnetic field at
the mG level. The subsequent synchrotron self-Compton emission expected demonstrated
that for such a magnetic field strength level, large inverse Compton emission is expected. 
Current gamma-ray sensitivities are shown to be able to constrain 
source properties like the magnetic field strength.

An alternative strong, $\mathcal{O}$(G), magnetic field scenario was also put forward, in which two
potential origins of the synchrotron emission are discussed. The first is that 
it is produced by electrons with a hard injection spectrum, which subsequently cool to
an $E^{-2}$ distribution. 
The second possibility is that this synchrotron emission is constantly produced 
by fresh electrons, picked up and accelerated near the edge of the remnant, as 
inferred for other similar astrophysical outflows.

For the strong magnetic field scenario considered, we also tested the efficiency of 
photo-meson production and photo-disintegration of CRs. The thermal component was demonstrated to be 
crucial for the interactions of the CRs in the outflow of the BNS merger remnant, especially
at early times, when it provides the dominant target radiation field for these processes. While for the assumed outflow speed the source 
is always optically thin to photomeson production, photo-disintegration of some nuclear isotopes is efficient in the first $\sim10$ days after the merger event.
Later, when the volume of the object has expanded, the thermal component becomes dimmer and the system is 
optically thin to all hadronic interactions. At this point, the maximum energy
of nuclei is no longer dictated by losses on this thermal radiation, allowing 
the acceleration up to energies beyond the ``ankle'', while protons can be accelerated up to the ankle.
In order for such a population to power the observed CR flux, a total energy output in CRs 
by the source of $\sim 10^{50}$~erg is required at late timescales.

An origin of CRs below the ``ankle'' related to NS merger remnants carries similarities 
with that put forward previously by others \cite{Anchordoqui:2007tn, Aloisio:2013hya, Globus:2014fka, Unger:2015laa,Aab:2016zth, Biehl:2017zlw}. Indeed, such a scenario rather naturally explains the apparent lightness of the CR composition inferred at these energies from elongation rate measurements by the Pierre Auger Observatory \cite{Aab:2017cgk}. 
The recent in-depth observations of such a remnant, over a broad energy range, have shed fresh light on the non-thermal aspects of this phenomenon, and continues to motivate them as promising CR sources.

Since the submission of this paper, a preprint of another work has been submitted, also considering CR acceleration in BNS merger remnants~\cite{Kimura:2018ggg}. The authors of this paper appear to have come to broadly similar conclusions to our own.

\section*{Acknowledgements}
We thank Martin Pohl, Walter Winter, and Stefan Ohm for useful discussions. 
This work has been supported by the European Research Council (ERC) under the European Union’s Horizon 2020 research and innovation programme (Grant No. 646623).

\bibliographystyle{elsarticle-num}


\begin{thebibliography}{10}
\expandafter\ifx\csname url\endcsname\relax
  \def\url#1{\texttt{#1}}\fi
\expandafter\ifx\csname urlprefix\endcsname\relax\def\urlprefix{URL }\fi
\expandafter\ifx\csname href\endcsname\relax
  \def\href#1#2{#2} \def\path#1{#1}\fi

\bibitem{TheLIGOScientific:2017qsa}
B.~Abbott, et~al., {GW170817: Observation of Gravitational Waves from a Binary
  Neutron Star Inspiral}, Phys. Rev. Lett. 119~(16) (2017) 161101.
\newblock \href {http://arxiv.org/abs/1710.05832} {\path{arXiv:1710.05832}},
  \href {http://dx.doi.org/10.1103/PhysRevLett.119.161101}
  {\path{doi:10.1103/PhysRevLett.119.161101}}.

\bibitem{Monitor:2017mdv}
B.~P. Abbott, et~al., {Gravitational Waves and Gamma-rays from a Binary Neutron
  Star Merger: GW170817 and GRB 170817A}, Astrophys. J. 848~(2) (2017) L13.
\newblock \href {http://arxiv.org/abs/1710.05834} {\path{arXiv:1710.05834}},
  \href {http://dx.doi.org/10.3847/2041-8213/aa920c}
  {\path{doi:10.3847/2041-8213/aa920c}}.

\bibitem{Savchenko:2017ffs}
V.~Savchenko, et~al., {INTEGRAL Detection of the First Prompt Gamma-Ray Signal
  Coincident with the Gravitational-wave Event GW170817}, Astrophys. J. 848~(2)
  (2017) L15.
\newblock \href {http://arxiv.org/abs/1710.05449} {\path{arXiv:1710.05449}},
  \href {http://dx.doi.org/10.3847/2041-8213/aa8f94}
  {\path{doi:10.3847/2041-8213/aa8f94}}.

\bibitem{Arcavi:2017fuz}
I.~Arcavi, et~al., {Energetic eruptions leading to a peculiar hydrogen-rich
  explosion of a massive star} (2017).
\newblock \href {http://arxiv.org/abs/1711.02671} {\path{arXiv:1711.02671}},
  \href {http://dx.doi.org/10.1038/nature24030}
  {\path{doi:10.1038/nature24030}}.

\bibitem{Coulter:2017wya}
D.~A. Coulter, et~al., {Swope Supernova Survey 2017a (SSS17a), the Optical
  Counterpart to a Gravitational Wave Source}, Science[Science358,1556(2017)].
\newblock \href {http://arxiv.org/abs/1710.05452} {\path{arXiv:1710.05452}},
  \href {http://dx.doi.org/10.1126/science.aap9811}
  {\path{doi:10.1126/science.aap9811}}.

\bibitem{Lipunov:2017dwd}
V.~M. Lipunov, et~al., {MASTER Optical Detection of the First LIGO/Virgo
  Neutron Star Binary Merger GW170817}, Astrophys. J. 850~(1) (2017) L1.
\newblock \href {http://arxiv.org/abs/1710.05461} {\path{arXiv:1710.05461}},
  \href {http://dx.doi.org/10.3847/2041-8213/aa92c0}
  {\path{doi:10.3847/2041-8213/aa92c0}}.

\bibitem{Soares-Santos:2017lru}
M.~Soares-Santos, et~al., {The Electromagnetic Counterpart of the Binary
  Neutron Star Merger LIGO/Virgo GW170817. I. Discovery of the Optical
  Counterpart Using the Dark Energy Camera}, Astrophys. J. 848~(2) (2017) L16.
\newblock \href {http://arxiv.org/abs/1710.05459} {\path{arXiv:1710.05459}},
  \href {http://dx.doi.org/10.3847/2041-8213/aa9059}
  {\path{doi:10.3847/2041-8213/aa9059}}.

\bibitem{Valenti:2017ngx}
S.~Valenti, D.~J. Sand, S.~Yang, E.~Cappellaro, L.~Tartaglia, A.~Corsi, S.~W.
  Jha, D.~E. Reichart, J.~Haislip, V.~Kouprianov, {The discovery of the
  electromagnetic counterpart of GW170817: kilonova AT 2017gfo/DLT17ck},
  Astrophys. J. 848~(2) (2017) L24.
\newblock \href {http://arxiv.org/abs/1710.05854} {\path{arXiv:1710.05854}},
  \href {http://dx.doi.org/10.3847/2041-8213/aa8edf}
  {\path{doi:10.3847/2041-8213/aa8edf}}.

\bibitem{Hjorth:2017yza}
J.~Hjorth, A.~J. Levan, N.~R. Tanvir, J.~D. Lyman, R.~Wojtak, S.~L. Schrøder,
  I.~Mandel, C.~Gall, S.~H. Bruun, {The Distance to NGC 4993: The Host Galaxy
  of the Gravitational-wave Event GW170817}, Astrophys. J. 848~(2) (2017) L31.
\newblock \href {http://arxiv.org/abs/1710.05856} {\path{arXiv:1710.05856}},
  \href {http://dx.doi.org/10.3847/2041-8213/aa9110}
  {\path{doi:10.3847/2041-8213/aa9110}}.

\bibitem{Hallinan:2017woc}
G.~Hallinan, et~al., {A Radio Counterpart to a Neutron Star Merger}, Science
  358 (2017) 1579.
\newblock \href {http://arxiv.org/abs/1710.05435} {\path{arXiv:1710.05435}},
  \href {http://dx.doi.org/10.1126/science.aap9855}
  {\path{doi:10.1126/science.aap9855}}.

\bibitem{Alexander:2017aly}
K.~D. Alexander, et~al., {The Electromagnetic Counterpart of the Binary Neutron
  Star Merger LIGO/VIRGO GW170817. VI. Radio Constraints on a Relativistic Jet
  and Predictions for Late-Time Emission from the Kilonova Ejecta}, Astrophys.
  J. 848~(2) (2017) L21.
\newblock \href {http://arxiv.org/abs/1710.05457} {\path{arXiv:1710.05457}},
  \href {http://dx.doi.org/10.3847/2041-8213/aa905d}
  {\path{doi:10.3847/2041-8213/aa905d}}.

\bibitem{Margutti:2017cjl}
R.~Margutti, et~al., {The Electromagnetic Counterpart of the Binary Neutron
  Star Merger LIGO/VIRGO GW170817. V. Rising X-ray Emission from an Off-Axis
  Jet}, Astrophys. J. 848~(2) (2017) L20.
\newblock \href {http://arxiv.org/abs/1710.05431} {\path{arXiv:1710.05431}},
  \href {http://dx.doi.org/10.3847/2041-8213/aa9057}
  {\path{doi:10.3847/2041-8213/aa9057}}.

\bibitem{Ruan:2017bha}
J.~J. Ruan, M.~Nynka, D.~Haggard, V.~Kalogera, P.~Evans, {Brightening X-Ray
  Emission from GW170817/GRB 170817A: Further Evidence for an Outflow},
  Astrophys. J. 853~(1) (2018) L4.
\newblock \href {http://arxiv.org/abs/1712.02809} {\path{arXiv:1712.02809}},
  \href {http://dx.doi.org/10.3847/2041-8213/aaa4f3}
  {\path{doi:10.3847/2041-8213/aaa4f3}}.

\bibitem{DAvanzo:2018zyz}
P.~D'Avanzo, et~al., {The evolution of the X-ray afterglow emission of GW
  170817 / GRB 170817A in XMM-Newton observations}, Astron. Astrophys. 613
  (2018) L1.
\newblock \href {http://arxiv.org/abs/1801.06164} {\path{arXiv:1801.06164}},
  \href {http://dx.doi.org/10.1051/0004-6361/201832664}
  {\path{doi:10.1051/0004-6361/201832664}}.

\bibitem{Nynka:2018vup}
M.~Nynka, J.~J. Ruan, D.~Haggard, {Fading of the X-ray Afterglow of Neutron
  Star Merger GW170817/GRB170817A at 260 days} (2018).
\newblock \href {http://arxiv.org/abs/1805.04093} {\path{arXiv:1805.04093}}.

\bibitem{Piro:2017ayh}
A.~L. Piro, J.~A. Kollmeier, {Evidence for Cocoon Emission from the Early Light
  Curve of SSS17a}, Astrophys. J. 855~(2) (2018) 103.
\newblock \href {http://arxiv.org/abs/1710.05822} {\path{arXiv:1710.05822}},
  \href {http://dx.doi.org/10.3847/1538-4357/aaaab3}
  {\path{doi:10.3847/1538-4357/aaaab3}}.

\bibitem{Takami:2013rza}
H.~Takami, K.~Kyutoku, K.~Ioka, {High-Energy Radiation from Remnants of Neutron
  Star Binary Mergers}, Phys. Rev. D89~(6) (2014) 063006.
\newblock \href {http://arxiv.org/abs/1307.6805} {\path{arXiv:1307.6805}},
  \href {http://dx.doi.org/10.1103/PhysRevD.89.063006}
  {\path{doi:10.1103/PhysRevD.89.063006}}.

\bibitem{Bell:2017zzx}
A.~R. Bell, A.~T. Araudo, J.~H. Matthews, K.~M. Blundell, {Cosmic Ray
  Acceleration by Relativistic Shocks: Limits and Estimates}, Mon. Not. Roy.
  Astron. Soc. 473~(2) (2018) 2364--2371.
\newblock \href {http://arxiv.org/abs/1709.07793} {\path{arXiv:1709.07793}},
  \href {http://dx.doi.org/10.1093/mnras/stx2485}
  {\path{doi:10.1093/mnras/stx2485}}.

\bibitem{Mooley:2017enz}
K.~P. Mooley, et~al., {A mildly relativistic wide-angle outflow in the neutron
  star merger GW170817}, Nature 554 (2018) 207.
\newblock \href {http://arxiv.org/abs/1711.11573} {\path{arXiv:1711.11573}},
  \href {http://dx.doi.org/10.1038/nature25452}
  {\path{doi:10.1038/nature25452}}.

\bibitem{hillas2004a}
A.~M. {Hillas}, {Where do 10$^{19}$ eV cosmic rays come from?}, Nuclear Physics
  B Proceedings Supplements 136 (2004) 139--146.
\newblock \href {http://dx.doi.org/10.1016/j.nuclphysbps.2004.10.004}
  {\path{doi:10.1016/j.nuclphysbps.2004.10.004}}.

\bibitem{Apel:2011mi}
W.~D. Apel, et~al., {Kneelike structure in the spectrum of the heavy component
  of cosmic rays observed with KASCADE-Grande}, Phys. Rev. Lett. 107 (2011)
  171104.
\newblock \href {http://arxiv.org/abs/1107.5885} {\path{arXiv:1107.5885}},
  \href {http://dx.doi.org/10.1103/PhysRevLett.107.171104}
  {\path{doi:10.1103/PhysRevLett.107.171104}}.

\bibitem{Giacinti:2011ww}
G.~Giacinti, M.~Kachelriess, D.~V. Semikoz, G.~Sigl, {Cosmic Ray Anisotropy as
  Signature for the Transition from Galactic to Extragalactic Cosmic Rays},
  JCAP 1207 (2012) 031.
\newblock \href {http://arxiv.org/abs/1112.5599} {\path{arXiv:1112.5599}},
  \href {http://dx.doi.org/10.1088/1475-7516/2012/07/031}
  {\path{doi:10.1088/1475-7516/2012/07/031}}.

\bibitem{Abreu:2012ybu}
P.~Abreu, et~al., {Constraints on the origin of cosmic rays above $10^{18}$ eV
  from large scale anisotropy searches in data of the Pierre Auger
  Observatory}, Astrophys. J. 762 (2012) L13.
\newblock \href {http://arxiv.org/abs/1212.3083} {\path{arXiv:1212.3083}},
  \href {http://dx.doi.org/10.1088/2041-8205/762/1/L13}
  {\path{doi:10.1088/2041-8205/762/1/L13}}.

\bibitem{Burbidge:1959}
G.~R. Burbidge, {Estimates of the total energy in particles and magnetic field
  in the non-thermal radio source}, Astrophys. J. 129.
\newblock \href {http://dx.doi.org/10.1086/146680} {\path{doi:10.1086/146680}}.

\bibitem{bell2004}
A.~R. {Bell}, {Turbulent amplification of magnetic field and diffusive shock
  acceleration of cosmic rays}, "Mon. Not. Roy. Astron. Soc." 353~(2) (2004)
  550--558.
\newblock \href {http://dx.doi.org/10.1111/j.1365-2966.2004.08097.x}
  {\path{doi:10.1111/j.1365-2966.2004.08097.x}}.

\bibitem{Jones1994}
F.~C. {Jones}, {A theoretical review of diffusive shock acceleration},
  "Astrophys. J." 90 (1994) 561--565.
\newblock \href {http://dx.doi.org/10.1086/191875} {\path{doi:10.1086/191875}}.

\bibitem{Fransson:1998ma}
C.~Fransson, C.-I. Bjornsson, {Radio emission and particle acceleration in sn
  1993j}, Astrophys. J. 509 (1998) 861.
\newblock \href {http://arxiv.org/abs/astro-ph/9807030}
  {\path{arXiv:astro-ph/9807030}}, \href {http://dx.doi.org/10.1086/306531}
  {\path{doi:10.1086/306531}}.

\bibitem{Tatischeff:2009kh}
V.~Tatischeff, {Radio emission and nonlinear diffusive shock acceleration of
  cosmic rays in the supernova SN 1993J}, Astron. Astrophys. 499 (2009) 191.
\newblock \href {http://arxiv.org/abs/0903.2944} {\path{arXiv:0903.2944}},
  \href {http://dx.doi.org/10.1051/0004-6361/200811511}
  {\path{doi:10.1051/0004-6361/200811511}}.

\bibitem{Aharonian:2000pv}
F.~A. Aharonian, {TeV gamma-rays from BL Lac objects due to synchrotron
  radiation of extremely high-energy protons}, New Astron. 5 (2000) 377--395.
\newblock \href {http://arxiv.org/abs/astro-ph/0003159}
  {\path{arXiv:astro-ph/0003159}}, \href
  {http://dx.doi.org/10.1016/S1384-1076(00)00039-7}
  {\path{doi:10.1016/S1384-1076(00)00039-7}}.

\bibitem{Fransson:2004kw}
C.~Fransson, C.-I. Björnsson, {Modeling the radio and x-ray emission of SN
  1993J and SN 2002ap}, Springer Proc. Phys. 99 (2005) 59--69.
\newblock \href {http://arxiv.org/abs/astro-ph/0404267}
  {\path{arXiv:astro-ph/0404267}}, \href
  {http://dx.doi.org/10.1007/3-540-26633-X\_8}
  {\path{doi:10.1007/3-540-26633-X\_8}}.

\bibitem{Troja:2017nqp}
E.~Troja, et~al., {The X-ray counterpart to the gravitational wave event GW
  170817}, Nature 551 (2017) 71--74, [Nature551,71(2017)].
\newblock \href {http://arxiv.org/abs/1710.05433} {\path{arXiv:1710.05433}},
  \href {http://dx.doi.org/10.1038/nature24290}
  {\path{doi:10.1038/nature24290}}.

\bibitem{Villar:2018mqw}
V.~A. Villar, et~al., {Spitzer Space Telescope Infrared Observations of the
  Binary Neutron Star Merger GW170817} (2018).
\newblock \href {http://arxiv.org/abs/1805.08192} {\path{arXiv:1805.08192}}.

\bibitem{Lyman:2018qjg}
J.~D. Lyman, et~al., {The optical afterglow of the short gamma-ray burst
  associated with GW170817}, Nat. Astron. 2~(9) (2018) 751--754.
\newblock \href {http://arxiv.org/abs/1801.02669} {\path{arXiv:1801.02669}},
  \href {http://dx.doi.org/10.1038/s41550-018-0511-3}
  {\path{doi:10.1038/s41550-018-0511-3}}.

\bibitem{Margutti:2018xqd}
R.~Margutti, et~al., {The Binary Neutron Star Event LIGO/Virgo GW170817 160
  Days after Merger: Synchrotron Emission across the Electromagnetic Spectrum},
  Astrophys. J. 856~(1) (2018) L18.
\newblock \href {http://arxiv.org/abs/1801.03531} {\path{arXiv:1801.03531}},
  \href {http://dx.doi.org/10.3847/2041-8213/aab2ad}
  {\path{doi:10.3847/2041-8213/aab2ad}}.

\bibitem{Abdalla:2017mtd}
H.~Abdalla, et~al., {TeV gamma-ray observations of the binary neutron star
  merger GW170817 with H.E.S.S}, Astrophys. J. 850~(2) (2017) L22.
\newblock \href {http://arxiv.org/abs/1710.05862} {\path{arXiv:1710.05862}},
  \href {http://dx.doi.org/10.3847/2041-8213/aa97d2}
  {\path{doi:10.3847/2041-8213/aa97d2}}.

\bibitem{Holler:2015uca}
M.~Holler, et~al., {Observations of the Crab Nebula with H.E.S.S. Phase II},
  PoS ICRC2015 (2016) 847.
\newblock \href {http://arxiv.org/abs/1509.02902} {\path{arXiv:1509.02902}},
  \href {http://dx.doi.org/10.22323/1.236.0847}
  {\path{doi:10.22323/1.236.0847}}.

\bibitem{Longair:1994wu}
M.~S. Longair, {High-energy astrophysics. Vol. 2: Stars, the galaxy and the
  interstellar medium}, Cambridge University Press, 1994.

\bibitem{Gould:1967zzb}
R.~J. Gould, G.~P. Schreder, {Pair Production in Photon-Photon Collisions},
  Phys. Rev. 155 (1967) 1404--1407.
\newblock \href {http://dx.doi.org/10.1103/PhysRev.155.1404}
  {\path{doi:10.1103/PhysRev.155.1404}}.

\bibitem{Neronov:2007mh}
A.~Y. Neronov, D.~V. Semikoz, I.~I. Tkachev, {Ultra-High Energy Cosmic Ray
  production in the polar cap regions of black hole magnetospheres}, New J.
  Phys. 11 (2009) 065015.
\newblock \href {http://arxiv.org/abs/0712.1737} {\path{arXiv:0712.1737}},
  \href {http://dx.doi.org/10.1088/1367-2630/11/6/065015}
  {\path{doi:10.1088/1367-2630/11/6/065015}}.

\bibitem{Murase:2010va}
K.~Murase, J.~F. Beacom, {Very-High-Energy Gamma-Ray Signal from Nuclear
  Photodisintegration as a Probe of Extragalactic Sources of Ultrahigh-Energy
  Nuclei}, Phys. Rev. D82 (2010) 043008.
\newblock \href {http://arxiv.org/abs/1002.3980} {\path{arXiv:1002.3980}},
  \href {http://dx.doi.org/10.1103/PhysRevD.82.043008}
  {\path{doi:10.1103/PhysRevD.82.043008}}.

\bibitem{Aharonian:2010te}
F.~Aharonian, A.~M. Taylor, {Limitations on the Photo-disintegration Process as
  a Source of VHE Photons}, Astropart. Phys. 34 (2010) 258--266.
\newblock \href {http://arxiv.org/abs/1005.3230} {\path{arXiv:1005.3230}},
  \href {http://dx.doi.org/10.1016/j.astropartphys.2010.08.004}
  {\path{doi:10.1016/j.astropartphys.2010.08.004}}.

\bibitem{Biehl:2017zlw}
D.~Biehl, D.~Boncioli, A.~Fedynitch, W.~Winter, {Cosmic-Ray and Neutrino
  Emission from Gamma-Ray Bursts with a Nuclear Cascade}, Astron. Astrophys.
  611 (2018) A101.
\newblock \href {http://arxiv.org/abs/1705.08909} {\path{arXiv:1705.08909}},
  \href {http://dx.doi.org/10.1051/0004-6361/201731337}
  {\path{doi:10.1051/0004-6361/201731337}}.

\bibitem{Rodrigues:2017fmu}
X.~Rodrigues, A.~Fedynitch, S.~Gao, D.~Boncioli, W.~Winter, {Neutrinos and
  Ultra-High-Energy Cosmic-Ray Nuclei from Blazars}, Astrophys. J. 854~(1)
  (2018) 54.
\newblock \href {http://arxiv.org/abs/1711.02091} {\path{arXiv:1711.02091}},
  \href {http://dx.doi.org/10.3847/1538-4357/aaa7ee}
  {\path{doi:10.3847/1538-4357/aaa7ee}}.

\bibitem{Gaisser:2013bla}
T.~K. Gaisser, T.~Stanev, S.~Tilav, {Cosmic Ray Energy Spectrum from
  Measurements of Air Showers}, Front. Phys.(Beijing) 8 (2013) 748--758.
\newblock \href {http://arxiv.org/abs/1303.3565} {\path{arXiv:1303.3565}},
  \href {http://dx.doi.org/10.1007/s11467-013-0319-7}
  {\path{doi:10.1007/s11467-013-0319-7}}.

\bibitem{Waxman:1995dg}
E.~Waxman, {Cosmological origin for cosmic rays above 10**19-eV}, Astrophys. J.
  452 (1995) L1--L4.
\newblock \href {http://arxiv.org/abs/astro-ph/9508037}
  {\path{arXiv:astro-ph/9508037}}, \href {http://dx.doi.org/10.1086/309715}
  {\path{doi:10.1086/309715}}.

\bibitem{ANTARES:2017bia}
A.~Albert, et~al., {Search for High-energy Neutrinos from Binary Neutron Star
  Merger GW170817 with ANTARES, IceCube, and the Pierre Auger Observatory},
  Astrophys. J. 850~(2) (2017) L35.
\newblock \href {http://arxiv.org/abs/1710.05839} {\path{arXiv:1710.05839}},
  \href {http://dx.doi.org/10.3847/2041-8213/aa9aed}
  {\path{doi:10.3847/2041-8213/aa9aed}}.

\bibitem{Fang:2017tla}
K.~Fang, B.~D. Metzger, {High-Energy Neutrinos from Millisecond Magnetars
  formed from the Merger of Binary Neutron Stars}, Astrophys. J. 849~(2) (2017)
  153, [Astrophys. J.849,153(2017)].
\newblock \href {http://arxiv.org/abs/1707.04263} {\path{arXiv:1707.04263}},
  \href {http://dx.doi.org/10.3847/1538-4357/aa8b6a}
  {\path{doi:10.3847/1538-4357/aa8b6a}}.

\bibitem{Kimura:2017kan}
S.~S. Kimura, K.~Murase, P.~Mészáros, K.~Kiuchi, {High-Energy Neutrino
  Emission from Short Gamma-Ray Bursts: Prospects for Coincident Detection with
  Gravitational Waves}, Astrophys. J. 848~(1) (2017) L4.
\newblock \href {http://arxiv.org/abs/1708.07075} {\path{arXiv:1708.07075}},
  \href {http://dx.doi.org/10.3847/2041-8213/aa8d14}
  {\path{doi:10.3847/2041-8213/aa8d14}}.

\bibitem{Anchordoqui:2007tn}
L.~A. Anchordoqui, D.~Hooper, S.~Sarkar, A.~M. Taylor, {High-energy neutrinos
  from astrophysical accelerators of cosmic ray nuclei}, Astropart. Phys. 29
  (2008) 1--13.
\newblock \href {http://arxiv.org/abs/astro-ph/0703001}
  {\path{arXiv:astro-ph/0703001}}, \href
  {http://dx.doi.org/10.1016/j.astropartphys.2007.10.006}
  {\path{doi:10.1016/j.astropartphys.2007.10.006}}.

\bibitem{Aloisio:2013hya}
R.~Aloisio, V.~Berezinsky, P.~Blasi, {Ultra high energy cosmic rays:
  implications of Auger data for source spectra and chemical composition}, JCAP
  1410~(10) (2014) 020.
\newblock \href {http://arxiv.org/abs/1312.7459} {\path{arXiv:1312.7459}},
  \href {http://dx.doi.org/10.1088/1475-7516/2014/10/020}
  {\path{doi:10.1088/1475-7516/2014/10/020}}.

\bibitem{Globus:2014fka}
N.~Globus, D.~Allard, R.~Mochkovitch, E.~Parizot, {UHECR acceleration at GRB
  internal shocks}, Mon. Not. Roy. Astron. Soc. 451~(1) (2015) 751--790.
\newblock \href {http://arxiv.org/abs/1409.1271} {\path{arXiv:1409.1271}},
  \href {http://dx.doi.org/10.1093/mnras/stv893}
  {\path{doi:10.1093/mnras/stv893}}.

\bibitem{Unger:2015laa}
M.~Unger, G.~R. Farrar, L.~A. Anchordoqui, {Origin of the ankle in the
  ultrahigh energy cosmic ray spectrum, and of the extragalactic protons below
  it}, Phys. Rev. D92~(12) (2015) 123001.
\newblock \href {http://arxiv.org/abs/1505.02153} {\path{arXiv:1505.02153}},
  \href {http://dx.doi.org/10.1103/PhysRevD.92.123001}
  {\path{doi:10.1103/PhysRevD.92.123001}}.

\bibitem{Aab:2016zth}
A.~Aab, et~al., {Combined fit of spectrum and composition data as measured by
  the Pierre Auger Observatory}, JCAP 1704~(04) (2017) 038, [Erratum:
  JCAP1803,no.03,E02(2018)].
\newblock \href {http://arxiv.org/abs/1612.07155} {\path{arXiv:1612.07155}},
  \href {http://dx.doi.org/10.1088/1475-7516/2018/03/E02,
  10.1088/1475-7516/2017/04/038} {\path{doi:10.1088/1475-7516/2018/03/E02,
  10.1088/1475-7516/2017/04/038}}.

\bibitem{Aab:2017cgk}
A.~Aab, et~al., {Inferences on mass composition and tests of hadronic
  interactions from 0.3 to 100 EeV using the water-Cherenkov detectors of the
  Pierre Auger Observatory}, Phys. Rev. D96~(12) (2017) 122003.
\newblock \href {http://arxiv.org/abs/1710.07249} {\path{arXiv:1710.07249}},
  \href {http://dx.doi.org/10.1103/PhysRevD.96.122003}
  {\path{doi:10.1103/PhysRevD.96.122003}}.

\bibitem{Kimura:2018ggg}
S.~S. Kimura, K.~Murase, P.~Mészáros, {Super-Knee Cosmic Rays from Galactic
  Neutron Star Merger Remnants}\href {http://arxiv.org/abs/1807.03290}
  {\path{arXiv:1807.03290}}.

\end{thebibliography}

\end{document}